# Decoupling edge versus bulk conductance in the trivial regime of an InAs/GaSb double quantum well using Corbino ring geometry

Binh-Minh Nguyen*, Andrey A. Kiselev, Ramsey Noah, Wei Yi, Fanming Qu, Arjan J.A. Beukman, Folkert K. de Vries, Jasper van Veen, Stevan Nadj-Perge, Leo P. Kouwenhoven, Morten Kjaergaard, Henri J. Suominen, Fabrizio Nichele, Charles M. Marcus, Michael J. Manfra, and Marko Sokolich*

HRL Laboratories, 3011 Malibu Canyon Rd, Malibu, CA 90265, USA

QuTech and Kavli Institute of Nanoscience, Delft University of Technology, 2600 GA Delft, Netherlands

Center for Quantum Devices, Niels Bohr Institute, University of Copenhagen, 2100 Copenhagen, Denmark

Department of Physics and Astronomy, and Station Q Purdue, Purdue University, West Lafayette, Indiana 47907, USA

**Abstract:**

A Corbino ring geometry is utilized to analyze edge and bulk conductance of InAs/GaSb quantum well structures. We show that edge conductance exists in the trivial regime of this theoretically-predicted topological system with a temperature insensitive linear resistivity per unit length in the range of 2 kΩ/μm. A resistor network model of the device is developed to decouple the edge conductance from the bulk conductance, providing a quantitative technique to further investigate the nature of this trivial edge conductance, conclusively identified here as being of *n*-type.

*Corresponding authors: mbnguyen@hrl.com and MSokolich@hrl.com

InAs/GaSb double quantum well structures have been long known to exhibit semi-metallic behavior when the electron energy level in the InAs well lies below the hole energy level in the GaSb well.[1-6] In such an "inverted" regime, the material system was predicted to be a topological quantum spin Hall insulator with insulating bulk and conducting helical edge states.[7] This theoretical proposal opened a new prospect for the double quantum well structure to be used in Majorana fermion devices and quantum computing.[8] Yet, it remains a great challenge to reliably identify the inverted regime and reveal its helical edge states.

As the electronic structure of the InAs/GaSb quantum wells can be continuously tuned under the electrostatic action of top and bottom gates, the most straightforward way to identify the inverted regime is to construct a 2D phase diagram as a function of top and back gate biases similar to what has been theoretically predicted.[7] Shown as an example in Figure 1 (a) is the phase diagram of the InAs/GaSb device as a function of back gate bias ($V_{BG}$) and top gate bias ($V_{TG}$) calculated using the capacitor model that we have introduced in Ref. 9 (see, specifically, Supplemental Material, Sec. III for details). Ideally, one would expect to be able to navigate in the gate bias space between the topological regime where electrons and holes coexist [region I in Figure 1 (a)] and the trivial regime where both type of carriers are depleted [region II in Figure 1 (a)]. However, in many cases, the weak and leaky backgate action due to a thick, defective buffer layer limits the range of operation of the back gate, thus hindering the ability to construct this 2D phase diagram. Without a complete 2D phase diagram, the inverted regime can only be indirectly inferred, e.g., based on the resistance behavior under in-plane[3,10] or out of plane[10,11] magnetic field. It was not until recently that the complete 2D phase diagram was experimentally demonstrated[9] with the use of high mobility materials and efficient back gate coupling[12].

To date, multiple approaches have been utilized to capture the signature of edge conductance in InAs/GaSb quantum well devices. The most common technique is based on non-local measurement of a Hall bar device which shows a non-local resistance, sometimes close to the theoretically predicted quantized resistance values.[13-18] Edge mode transport was also deduced from superconducting quantum interference patterns in a superconductor-InAs/GaSb-superconductor Josephson junction device.[19] Magnetic flux image reconstruction employing a superconductor quantum interference device (SQUID) can also visualize edge conductance in an InAs/GaSb Hall bar.[20] Nevertheless, Refs. [13-20] are lacking in providing clear evidence that these reported edge transports were observed in the inverted regime. In lieu, given the existence of the edge conductance, it is taken for granted that the device is in the inverted regime. However, the observed RP was found to be insensitive to in-plane magnetic field,[16,19] which contradicts the inverted regime identification criterion[3,10] and requires further theoretical hypotheses[21]. Very recently, Nichele et al.[22] reported that edge conductance is present in the trivial regime, as observed in both transport and scanning SQUID microscopy. In this report, we further substantiate this claim with analysis of a novel Corbino ring device, confirming the presence of edge channels in the trivial regime. In addition, the edge and bulk resistivities have been extracted and mapped in a two dimensional gate bias space of the top and back gates, revealing the *n*-type nature of the edge conductance in the trivial regime.

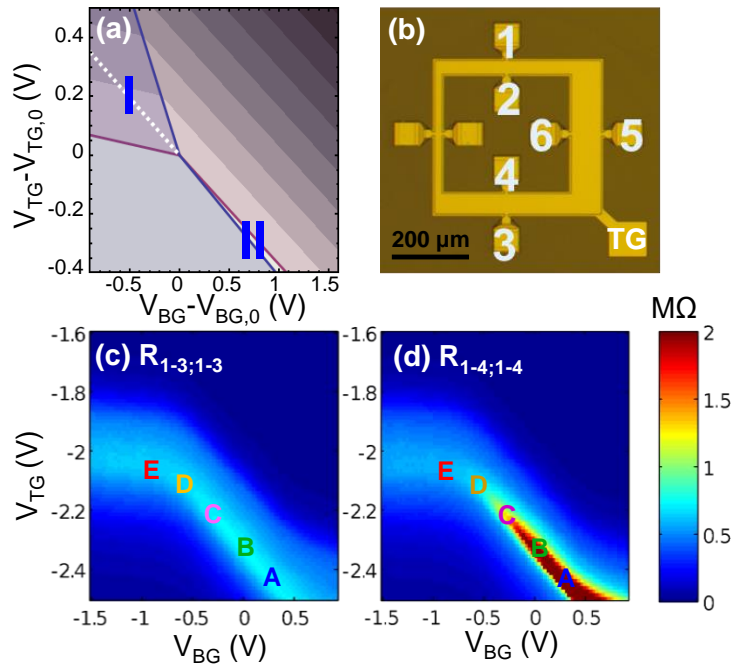

*Figure 1 (a) Simulated phase diagram (based on actual device geometries — InAs-on-GaSb heterostack reported in Ref. [12] with a 40 nm HfOx top gate dielectric) illustrating the position of two regimes: inverted CNP (region I — dashed line) and trivial resistance peak (region II with narrow white sector); The contour lines in the electron-rich sectors mark constant electron densities at multiples of $9.66 \times 10^{10}$ cm$^{-2}$, corresponding, at B=2 T, to filling of complete Landau levels (b) optical micrograph of a Corbino ring device with top gate lead; Resistance maps at 1.8K for (c) with-edge $R_{1-3;1-3}$ and (d) no-edge $R_{1-4;1-4}$ (see text for a definition). Points A-E in Figure 1 (c), (d) indicate gate biases where temperature dependent measurements were performed (see Supplementary Material, Section II).*

In order to decouple the edge and bulk contributions, we utilized a Corbino-type device geometry allowing for a measurement setup without conductance paths attainable via an edge channel alone. In addition, we chose a Corbino ring geometry with multiple inner and outer pairs of contact leads [Figure 1 (b)] in order to isolate the contribution of contact resistance that is unavoidable in a solid disk Corbino device with only two terminals[16]. This allows for reliable measurements further outside of the RPs where the device resistance is extremely small and cannot be measured in a Corbino disk geometry. Indeed, in this multiple-lead configuration of Corbino ring devices, non-local measurement is possible, similar to multi-lead Hall bars, but with the added advantage of allowing both "with-edge" configurations, i.e., those containing conductance paths attainable via an edge channel alone, and "no-edge" ones (with no such paths) on the same device. This is critical for quantitative extraction of edge and bulk resistivity, especially at the RPs where the bulk resistance is very high compared to edge resistance. Transport at the RPs in Hall bar-type devices is dominated by edge transport,[19,20] making it hard to deduce the bulk contribution.[13-17] Whereas, for Corbino ring devices, the interplay between bulk and edge conductance can be clearly observed and extracted from the same device, which can eliminate processing uncertainty causing device-to-device variation. As will be detailed in this report, analysis of a Corbino ring device provides a simple yet effective method to extract the contribution of edge conductance, similar to what could be achieved by more elaborated SQUID measurements.

Devices used in this work were grown and fabricated using the same procedure as the previous study.[12] The only difference is that the wafer was grown in a different growth campaign, and resulted in slightly lower mobility, 200,000 cm²/Vs in this material, in comparison with 500,000 cm²/Vs in our previously reported results[12], both at $N_s=10^{12}$ cm$^{-2}$. It is important to note that this lower mobility is comparable to values reported by others[4,10,11] and that lowering mobility, by intentionally introducing disorder,[16,23] is generally accepted as a means to suppress bulk conductivity. Figure 1 (b) shows an optical micrograph of the Corbino ring device with inner and outer square size of 340 and 440 µm, respectively. Channel widths between pairs 1-2, 3-4 and 5-6 are 40, 60, 80 µm, respectively. The left pair of leads with 20 µm separation was not used in this study. All data were collected at 1.8K (except for the temperature dependent measurements presented in the Supplementary Material).

Figure 1 (c) and (d) show the 2D resistance maps for the with-edge $R_{1-3;1-3}$ and no-edge $R_{1-4;1-4}$, respectively. The notation $R_{i-j;k-l}$ indicates current fed through leads *i* and *j* and voltage recorded between leads *k* and *l*. Despite having nominally the same structural design, lower mobility wafers in this growth campaign consistently show only one branch of the phase diagram while the high mobility materials[12] exhibited a clear delineation between the inverted and the normal regimes (see Supplemental Material Figure S1 and Ref.9). Devices of limited tunability were also reported in Refs.[15,19] where edge conductance was shown to exist. The correlation between mobility and observation of the inverted regime in the 2D phase diagram is still under investigation, but is beyond the scope of this report, which highlights the presence of edge conductance in the trivial regime.

Comparing Figure 1 (c) and (d), a clear distinction between with-edge and no-edge peak resistances (e.g., *7x10⁵ Ω* for $R_{1-3;1-3}$ vs *4x10⁶ Ω* for $R_{1-4;1-4}$ at the same point A) is a strong evidence for edge conductance. This is not entirely unexpected as it is well known that the free surface of InAs creates an accumulation channel for electrons.[24-28] Since the no-edge resistance is a few times larger than the with-edge one, we conclude that $R_{1-3;1-3}$ is mostly due to the outer mesa edges of 880 μm length (with 2 edges in parallel), giving a rough estimate of 1.6 kΩ/μm for the linear edge resistivity per micron (Figure S 4). This number matches with the length normalization from inner path $R_{2-4;2-4}$ and a Hall bar geometry (Figure S 5). This is also in agreement with what we get from a more quantitative extraction, as will be shown later.

With the InAs-on-GaSb heterostack[12] used in this work, the back gate acts primarily as the hole gate, and the top gate acts as the electron gate. The bottom right corner of the phase diagram corresponds to the depletion of both electron (negative top gate) and hole (positive back gate), suggesting the RPs we are observing are those of the trivial regime for the measured gate bias range. The arrow-shape of the RP pointing toward the top left corner also indicates that the gap is closing, similar to what one would expect theoretically (Figure 1 (a) and Ref. 7 ). The gap closure is also confirmed by the temperature dependence of RP with/without edge, as shown in Figure S 2.

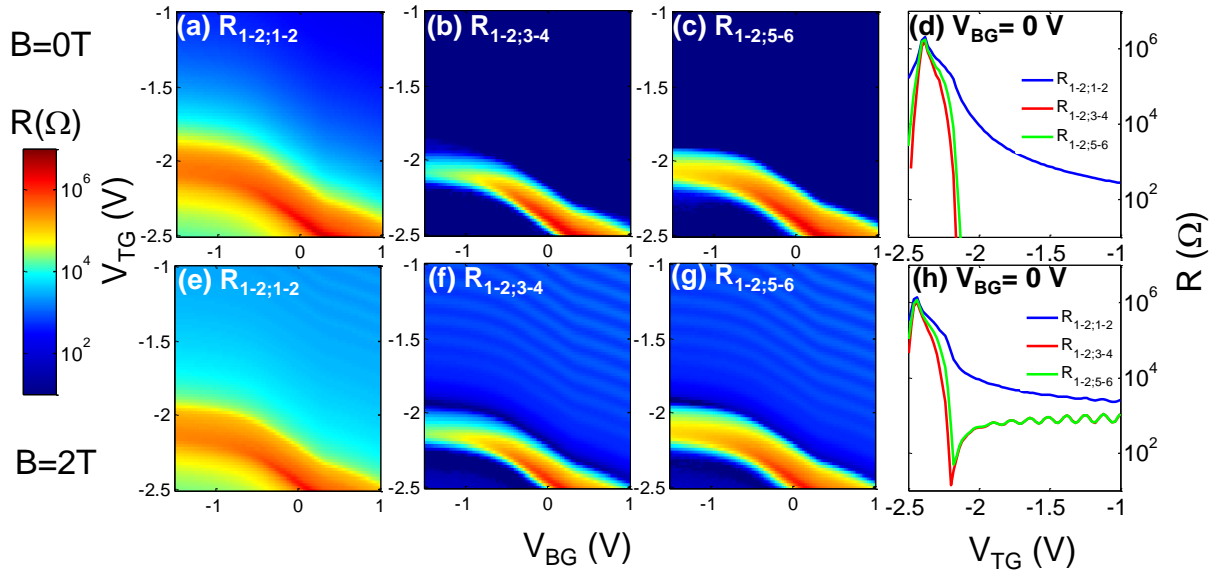

Figure 2. Local and non-local resistance maps (measured at 1.8K) at B= 0 T (a,b,c) and 2 T (e,f,g) indicating an edge conductance path at the RPs; (d) and (h) are 1D line cuts at $V_{BG}$= 0 V.

Evidence for edge conductance is reinforced with the local and non-local resistance measurement, as shown in Figure 2 and Supplementary Material, Section V. Despite the fact that the current is fed through one pair of opposite leads (through 1-2 leads in Figure 2, or through 3-4 or 5-6 in Figure S 6), at RPs, a substantial voltage drop is measured across the other pairs. This can only be explained if the edge conductance is significant, forming a quasi-equipotential line along the device edge. While the resistance at RPs peaks are similar, outside of the RP, the resistance in the electron regime (top right side of the RP ridge) is in the range of $10^3$ $\Omega$ for diagonal resistance $R_{1-2;1-2}$, $R_{3-4;3-4}$, $R_{5-6;5-6}$, but is in the noise floor (a few Ohms) for $R_{1-2;3-4}$ and $R_{1-2;5-6}$. This is due to the presence of non-negligible contact resistance (sub-k$\Omega$ range) when the channel resistance (few $\Omega$) is low. Even if contact resistance were completely eliminated, the low bulk resistance would make the current flow most directly between leads 1 and 2, resulting in a progressively lower voltage build up between the other (left floating) leads, (see Supplementary Material, Section III). This prevents quantitative analysis of edge vs bulk contribution further outside the RP. Nevertheless, using non-local measurements we are able to extend accessible dynamic range at least by an order of magnitude.

To highlight the superior dynamic range of this novel Corbino ring design over conventional Hall bars and Cobino disk, we apply a perpendicular magnetic field and document a formation of the integer quantum Hall effect (IQHE). Bulk resistivity is enhanced due to formation of the localized Landau orbitals. At $B=2$ T [Figure 2 (e)-(h), and $B=1$ T in supporting Figure S 8], the resistance map of $R_{1-2;1-2}$ does not seem to change much because it is still dominated by contact resistance outside of the RP, oscillations only appear in the top right corner with high electron density. However, in non-local measurement $R_{1-2;3-4}$ and $R_{1-2;5-6}$ where contact resistance is eliminated, with the resistance lifted from noise floor only to ~$10^2$ $\Omega$ range, the oscillations are clearly visible and appear much closer to the RP. It is worth noting that the oscillations in two separate measurements of $R_{1-2;3-4}$ and $R_{1-2;5-6}$ are perfectly in phase, indicating great measurement stability with no shift or drift with gate bias within one cooldown cycle. More interestingly, between the

RP and the electron-rich resistance plateau, there is a "canyon" of lower resistance, which cannot be observed in direct measurement of $R_{1-2;1-2}$ [Figure 2 (d), (h)]. This resistance canyon and plateau come from a crossover between disorder-limited bulk conduction and gradual formation of the IQHE regime as the device disorder is progressively screened by increasing carrier density. Indeed, close to the RP, behavior at finite magnetic field is similar to the $B=0$ case as material is too disordered for IQHE. Even though finite $B$ is kept unchanged, as carrier density increases, screening of disorder improves, leading to better localization of carriers by magnetic field, suppression of bulk conduction and formation of (semi-ballistic) chiral edge states. We stress that these IQHE protected edge states introduce a parallel edge conduction of a completely different nature (although both are likely facilitated by potential bending near edges). Ripplings in resistance maps manifest consecutive filling of disorder-broadened Landau levels (LL) in the bulk of the device, also accompanied by creation, one by one, of additional chiral edge channels. With electron density of $9.66 \times 10^{10}$ cm$^{-2}$ per single LL (when accounting for two spin subbands), and filling of up to about 10 individual LLs visible, Figure 2 (f) and (g) presents a very detailed 2D gate bias map of the electron density. Self-consistent Schrödinger-Poisson simulations[29] of the device stack, closely matching results of the equivalent capacitor model [see Figure 1 (a)], accurately reproduce this experimental bias-dependence of electron density requiring, e.g., at $V_{BG}=0.5V$, about 0.136V of additional top gate bias per LL versus 0.134V per LL measured.

To gain better understanding of the interplay between the edge and bulk conductances in multiport devices of complex geometries, a 2D resistor network was modelled, parametrized by linear edge and bulk resistivities: $\rho_{edge}$ ($\Omega/\mu m$) and $\rho_{bulk}$ ($\Omega/\square$) *(See Supplementary Material, Section III)*. Results of this numerical extraction are shown in Figure 3 both in absence of the magnetic field and at $B=2$ T (additional extracted maps for the case of $B=1$ T are shown in supporting Figure S 9).

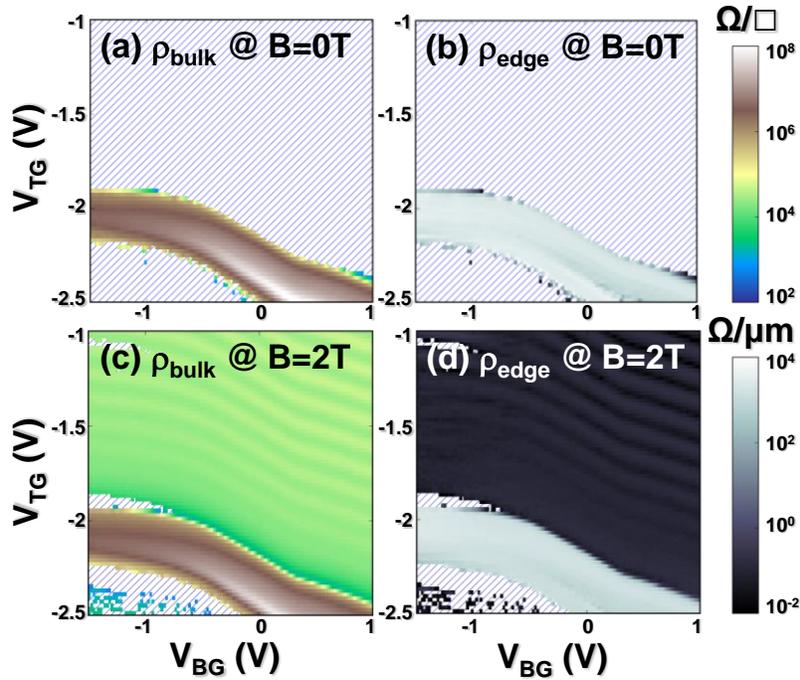

*Figure 3. 2D maps of extracted bulk and edge resistivities at 1.8K: (a), (b) at B=0 and (c), (d) at B=2 T.*

As expected, at *B=0 T*, the extraction procedure is meaningful only near the RPs. Further outside of the RPs, the channel resistance, governed by the diminishing bulk resistivity, becomes so small that voltage signals measured at the 3-4 and 5-6 leads are in the noise level, preventing further quantification of $\rho_{bulk}$ and $\rho_{edge}$ (beyond knowing that $\rho_{bulk}$ is smaller than its noise-level bound). In the gate bias domain allowing for quantitative extraction, the obtained $\rho_{bulk}$ is a sharp function of gate biases spanning more than 3 orders of magnitude, while $\rho_{edge}$ is relatively flat (all the way up to the domain edges).

Under a magnetic field, the enhanced bulk resistance outside the RPs (also accompanied by introduction of strong additional edge conduction of different nature) made the signal at the 3-4 and 5-6 leads measureable, thus facilitating the reliable numerical extraction over the whole 2D map. The crossover canyon at the RP boundary, mentioned above, is present again in the bulk resistivity map [Figure 3 (c)] along with a large "step" in $\rho_{edge}$ [Figure 3 (d)]. The extracted $\rho_{bulk}$ and $\rho_{edge}$ maps both show correlated ripplings due to LL filling, as expected: when Fermi level falls half-way between LLs, bulk turns most resistive while, simultaneously, IQHE edge channels are most protected and, consequently, most conductive. Indeed, the ripplings in $\rho_{bulk}$ and $\rho_{edge}$ are almost in anti-phase as can be best seen by examining the upper-right corner of two maps in Figure 3 (c) and (d).

The same extraction procedure was performed on data collected at different temperatures, revealing the decoupled temperature dependence for bulk and edge resistivities. Similar to the temperature dependent trend for resistances in Figure S 1 (a), (b), the bulk resistivity [Figure S 1 (d)] depends exponentially on temperature, and the energy gap is closing when going from A to E. The edge resistivity [Figure S 1 (d)] stays at ~2 kΩ/μm, which is consistent with the values extracted above from edge dominated resistance. This resistivity is insensitive to the measured temperatures between 1.8K and 25K. It is important to note that at this resistivity level, a mesoscopic Hall bar with lengths of a few micrometers can easily attain a resistance close to the theoretically predicted quantized conductance purely by coincidence. Thus measuring a non-local conductance near the expected quantized value is not necessarily a solid proof of helical edge transport in the inverted regime.

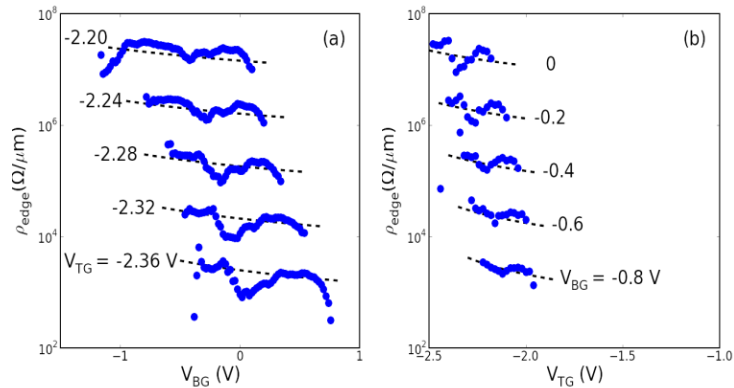

*Figure 4. Dots in (a) and (b) represent 1D line cuts along $V_{BG}$ and $V_{TG}$ directions, respectively, through the 2D map of the extracted edge resistivity [B=0 case, Figure 3 (b)]. The lowest cut in each subfigure is shown with its true value, while each consecutive cut is offset vertically by a factor of 10. All dashed lines collectively represent a simple three-parameter fit to the extracted data using an expansion of edge*

*conductivity $1/\rho_{edge}$ up to the first order in ($V_{BG}$, $V_{TG}$). Using this global fit, modulation of edge conductance across the domain of reliable extraction exceeds 2x.*

The ability to independently map the bulk and edge resistivities in a 2D top-back gate bias space allows for a further assessment on the nature of the edge channel. Shown in Figure 4 (a) and (b) are 1D line cuts of the edge resistivity as a function of back and top gate biases, respectively. In the regime where extraction is reliable (high bulk resistivity), the edge resistivity exhibits a negative slope with respect to both top and back gates. This signifies higher edge conductance with more positive gate bias, which is a behavior of *n*-type transport. The electron-like behavior of the edge conductance can be explained by the well-known Fermi level pinning and electron accumulation at the surface of InAs.[24-28] It is important to note that this trend for edge resistivity is resolved well beyond the numerical uncertainty of the extraction method. Indeed, the same trend can also be observed in the raw measurement resistance data for conductance paths that directly involve edge channels, like the $R_{1-3;1-3}$ of the Corbino ring (See Figure S 4), or longitudinal resistance of a Hall bar (See Figure S 5).

In summary, we have shown evidence that the trivial regime of InAs/GaSb double quantum well structures can host an *n*-type edge conductance channel. The Corbino ring geometry provided a simple device prototype that enables analysis of the interplay between the bulk and edge conduction channels, and, together with the resistor network model, offered a template for independent quantification of the bulk and edge conductivities. The ability to modulate the edge conductance with gate biases opens hope to controllably suppress the trivial edge conductance and reveal the helical topological edge states in future investigations.

Acknowledgements: This work was partially supported by Microsoft Station Q, the Netherlands Foundation for Fundamental Research on Matter (FOM), the Danish National Research Foundation, and the European Commission through the Marie Curie Fellowship Program.

**List of reference:**

1  Y. Naveh and B. Laikhtman,  Applied Physics Letters **66** (15), 1980 (1995).
2  S. de-Leon, L. D. Shvartsman, and B. Laikhtman,  Physical Review B **60** (3), 1861 (1999).
3  M. J. Yang, C. H. Yang, B. R. Bennett, and B. V. Shanabrook,  Physical Review Letters **78** (24), 4613 (1997).
4  M. Drndic, M. P. Grimshaw, L. J. Cooper, D. A. Ritchie, and N. K. Patel,  Applied Physics Letters **70** (4), 481 (1997).
5  L. J. Cooper, N. K. Patel, V. Drouot, E. H. Linfield, D. A. Ritchie, and M. Pepper,  Physical Review B **57** (19), 11915 (1998).
6  K. Suzuki, K. Takashina, S. Miyashita, and Y. Hirayama,  Physical Review Letters **93** (1), 016803 (2004).
7  C. Liu, T. L. Hughes, X.-L. Qi, K. Wang, and S.-C. Zhang,  Physical Review Letters **100** (23), 236601 (2008).
8  S. Mi, D. I. Pikulin, M. Wimmer, and C. W. J. Beenakker,  Physical Review B **87** (24), 241405 (2013).
9  F. Qu, A. J. A. Beukman, S. Nadj-Perge, M. Wimmer, B.-M. Nguyen, W. Yi, J. Thorp, M. Sokolich, A. A. Kiselev, M. J. Manfra, C. M. Marcus, and L. P. Kouwenhoven,  Physical Review Letters **115** (3), 036803 (2015).
10  I. Knez, PhD thesis, Rice University, 2012.
11  F. Nichele, A. N. Pal, P. Pietsch, T. Ihn, K. Ensslin, C. Charpentier, and W. Wegscheider,  Physical Review Letters **112** (3), 036802 (2014).
12  B.-M. Nguyen, W. Yi, R. Noah, J. Thorp, and M. Sokolich,  Applied Physics Letters **106** (3), 032107 (2015).
13  I. Knez, R.-R. Du, and G. Sullivan,  Physical Review Letters **107** (13), 136603 (2011).
14  K. Suzuki, Y. Harada, K. Onomitsu, and K. Muraki,  Physical Review B **87** (23), 235311 (2013).
15  I. Knez, C. T. Rettner, S.-H. Yang, S. S.-P. Parkin, L. Du, R.-R. Du, and G. Sullivan,  Physical Review Letters **112** (2), 026602 (2014).
16  L. Du, I. Knez, G. Sullivan, and R.-R. Du,  Physical Review Letters **114** (9), 096802 (2015).
17  S. Mueller, A. N. Pal, M. Karalic, T. Tschirky, C. Charpentier, W. Wegscheider, K. Ensslin, and T. Ihn,  Physical Review B **92** (8), 081303 (2015).
18  K. Suzuki, Y. Harada, K. Onomitsu, and K. Muraki,  Physical Review B **91** (24), 245309 (2015).
19  V. S. Pribiag, A. J. A. Beukman, F. Qu, M. C. Cassidy, C. Charpentier, W. Wegscheider, and L. P. Kouwenhoven,  Nat Nano **10** (7), 593 (2015).
20  E. M. Spanton, K. C. Nowack, L. Du, G. Sullivan, R.-R. Du, and K. A. Moler,  Physical Review Letters **113** (2), 026804 (2014).
21  D. I. Pikulin and T. Hyart,  Physical Review Letters **112** (17), 176403 (2014).
22  F. Nichele, H. J. Suominen, M. Kjaergaard, C. M. Marcus, E. Sajadi, J. A. Folk, F. Qu, A. J. A. Beukman, F. K. d. Vries, J. v. Veen, S. Nadj-Perge, L. P. Kouwenhoven, B.-M. Nguyen, A. A. Kiselev, W. Yi, M. Sokolich, M. J. Manfra, E. M. Spanton, and K. A. Moler,  arXiv:1511.01728 (2015).
23  C. Charpentier, S. Fält, C. Reichl, F. Nichele, A. Nath Pal, P. Pietsch, T. Ihn, K. Ensslin, and W. Wegscheider,  Applied Physics Letters **103** (11) (2013).
24  C. A. Mead and W. G. Spitzer,  Physical Review Letters **10** (11), 471 (1963).
25  J. M. Woodall, J. L. Freeouf, G. D. Pettit, T. Jackson, and P. Kirchner,  Journal of Vacuum Science and Technology **19** (3), 626 (1981).
26  M. Noguchi, K. Hirakawa, and T. Ikoma,  Physical Review Letters **66** (17), 2243 (1991).

27 D. C. Tsui, Physical Review Letters **24** (7), 303 (1970).
28 L. Ö. Olsson, C. B. M. Andersson, M. C. Håkansson, J. Kanski, L. Ilver, and U. O. Karlsson, Physical Review Letters **76** (19), 3626 (1996).
29 W. Yi, A. A. Kiselev, J. Thorp, R. Noah, B.-M. Nguyen, S. Bui, R. D. Rajavel, T. Hussain, M. F. Gyure, P. Kratz, Q. Qian, M. J. Manfra, V. S. Pribiag, L. P. Kouwenhoven, C. M. Marcus, and M. Sokolich, Applied Physics Letters **106** (14), 142103 (2015).

# Supplementary Material

# Decoupling edge versus bulk conductance in the trivial regime of InAs/GaSb double quantum well using Corbino ring geometry.

Binh-Minh Nguyen*, Andrey A. Kiselev, Ramsey Noah, Wei Yi, Fanming Qu, Arjan J.A. Beukman, Folkert K. de Vries, Jasper van Veen, Stevan Nadj-Perge, Leo P. Kouwenhoven, Morten Kjaergaard, Henri J. Suominen, Fabrizio Nichele, Charles M. Marcus, Michael J Manfra, Marko Sokolich*

*HRL Laboratories, 3011 Malibu Canyon Rd, Malibu, CA 90265, USA*

*QuTech and Kavli Institute of Nanoscience, Delft University of Technology, 2600 GA Delft, Netherlands*

*Center for Quantum Devices, Niels Bohr Institute, University of Copenhagen, 2100 Copenhagen, Denmark*

*Department of Physics and Astronomy, and Station Q Purdue, Purdue University, West Lafayette, Indiana 47907, USA*

*Corresponding authors: mbnguyen@hrl.com and MSokolich@hrl.com

# Contents

# I. Evidence of trivial and inverted regimes in high mobility wafers

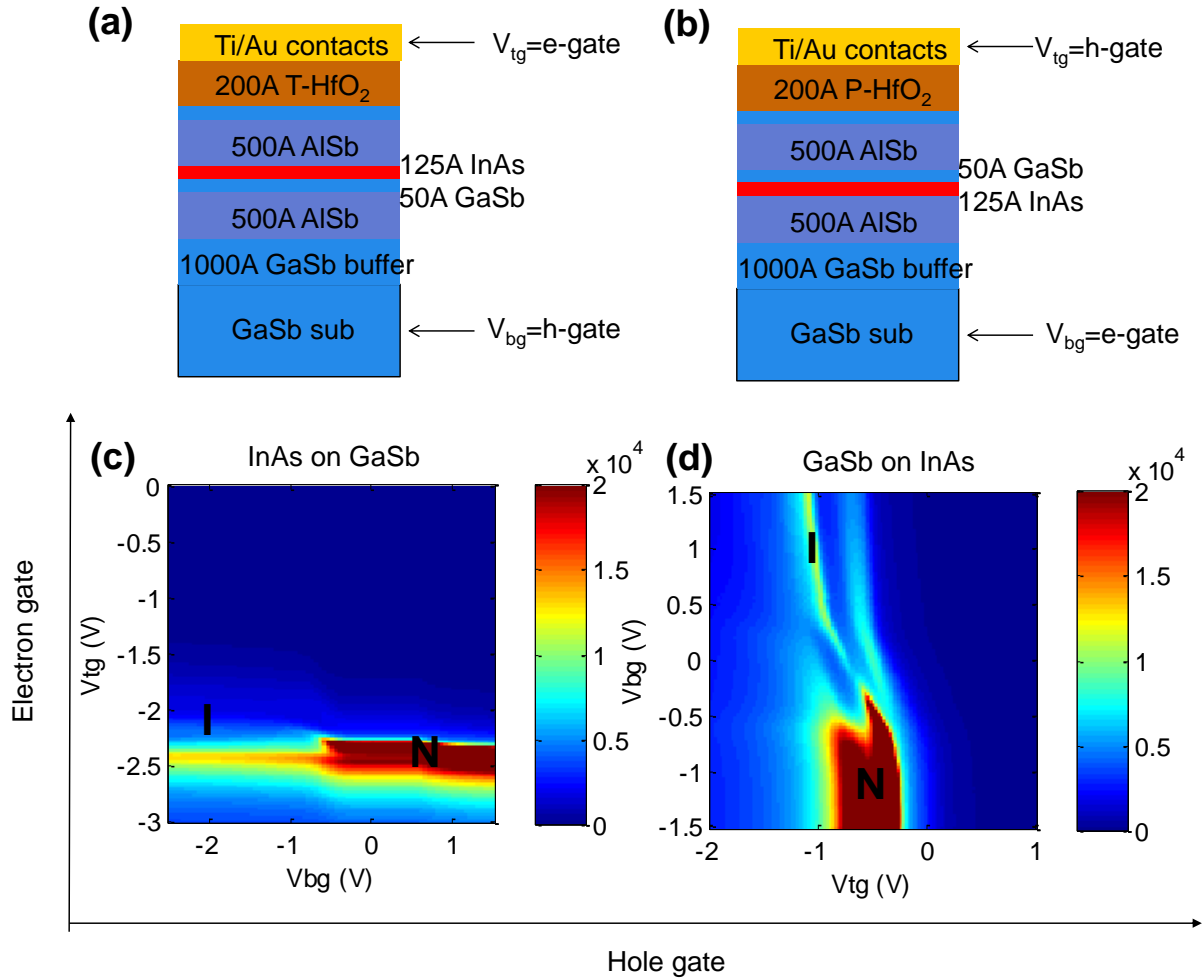

Figure S 1. *Device schematic and 2D phase diagram of high mobility wafers exhibiting a clear delineation between trivial and inverted regimes. (a,c) Wafer with InAs on GaSb sequencing similar to Ref 1 and Ref 2 and (b,d) Wafer with GaSb on InAs sequencing. All devices were processed using procedure reported in Ref. 1.*

## II. Temperature dependence measurement and extraction

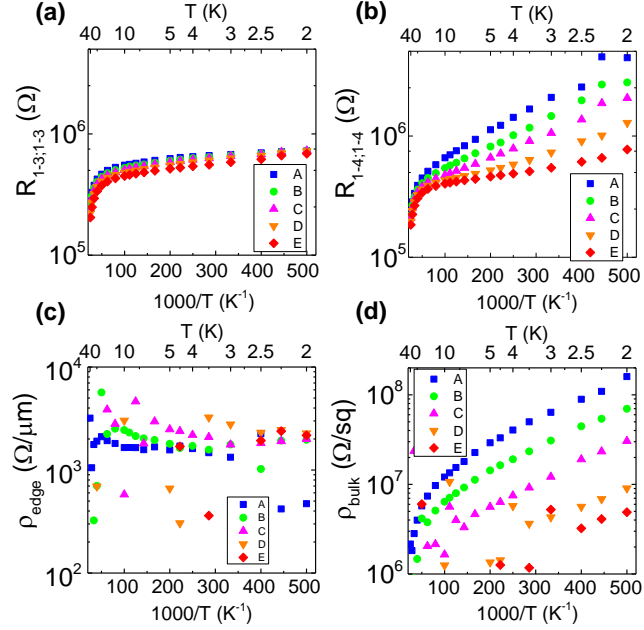

*Figure S 2. Temperature dependence of (a) $R_{1-3;1-3}$ and (b) $R_{1-4;1-4}$, along with extracted values for (c) edge resistivity and (d) bulk resistivity at points A-E in Figure 1 (c), (d), main text. Edge resistivity and edge-dominated transport are temperature insensitive while bulk resistivity and bulk-dominated transport exhibit an Arrhenius trend with temperature.*

As shown in Figure S 2 (a), (b), the temperature dependence of the RP peak resistance has two distinct regimes: below and above 10 K. Below 10 K, with edge contribution, the resistance is relatively insensitive to temperature while without edge, the resistance shows a clear dependence on temperature. Going from A to E, the trend is weakened, and eventually becomes temperature insensitive, which could be explained by a gapless scenario. Beyond point E, we did not see the gap reopening into the inverted regime, unlike previous high mobility devices (see Figure S 1 and Ref. *2*). The energy gap Δ was extracted from the Arrhenius trend $R \sim \exp(\Delta/2k_BT)$ to be ~13 K for point A and vanishing at E. At higher temperature regime (*T*>10 K), both resistances, with and without edges, roll over to a steeper slope, likely due to the fact that the bulk resistivity becomes so low that it now dominates the edge channel. The new energy gap is ~80 K, although reliable estimation is difficult. The two-gap trend is similar to what was reported in Ref.

[3] where the larger gap at high temperatures was assigned to be the hybridization-induced minigap and the smaller gap at low temperatures was assigned to be the localization gap. In Ref. [3], the transition between the two gap regimes also coincided with the turn on of the edge conductance, similar to what we see here in Figure S 2. However, in our case where the device was clearly in the trivial regime, the two energy gaps should not be due to the hybridization effect. Yet, its magnitude is conspicuously different from the expected direct gap between the conduction and valence states in the double layer. We speculate that a smaller gap could stem from disorder percolation[4] or from a localized shallow impurity level to either the conduction or valence band[5].

## III. Extraction of bulk and edge resistivities

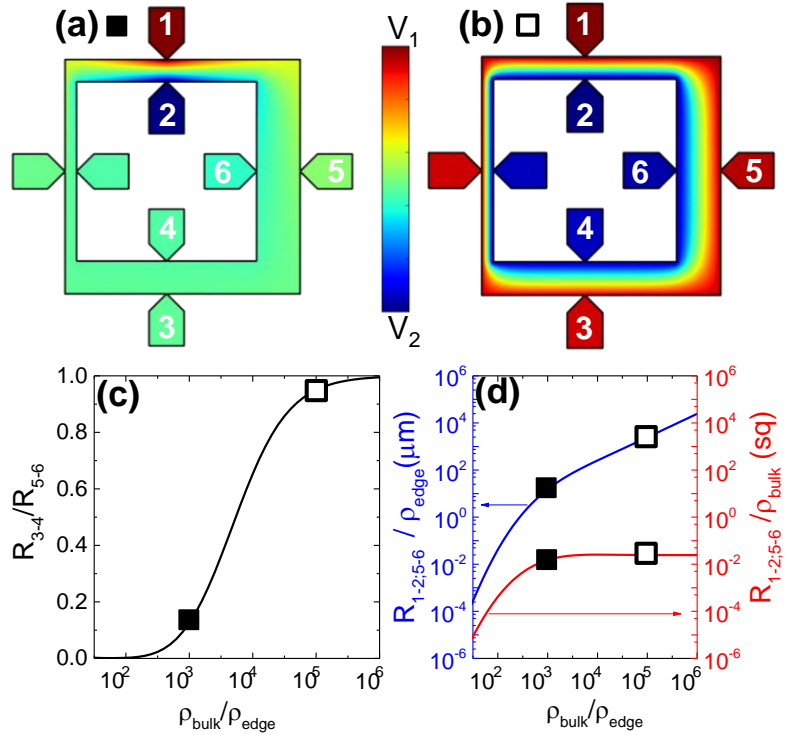

Figure S 3. Bias distribution maps simulated in the resistor network model for two cases of (a) modest ($10^3$) and (b) large ($10^5$) bulk-to-edge resistivity ratio; (c) resistance ratio $R_{1-2;3-4}/R_{1-2;5-6}$ and (d) scale factors $R_{1-2;5-6}/\rho_{edge}$ and $R_{1-2;5-6}/\rho_{bulk}$ as a function of bulk-to-edge resistivity ratio. Markers ■ and □ in (c),

(d) indicate the resistivity ratios used in the simulations depicted in (a) and (b), respectively. The arrows in (d) indicate the corresponding axis.

Shown in Figure S 3 (a) and (b) are bias distribution maps of our device when current is fed through terminals 1-2, for two representative $\rho_{bulk}/\rho_{edge}$ ratios of $10^3$ and $10^5$ (μm per □). Smaller resistivity ratio means the bulk is relatively more conductive, leading to the current running straight in the mesa between terminals 1 and 2, while with larger resistivity ratio, the bulk becomes relatively more insulating, the lower resistance edges form quasi-equipotential contours extending around the whole device. This transition is exactly what we have observed in the experiment.

Beyond qualitative explanation of experimental data, the resistor network simulation also allows for a simultaneous extraction of edge and bulk resistivities. As shown in Figure S 3 (c), the resistance ratio $R_{1-2;\,3-4}/R_{1-2;5-6}$ is dictated by the ratio of bulk over edge resistivity alone, so the resistivity ratio can be extracted from the measured resistance ratio $R_{1-2;\,3-4}/R_{1-2;5-6}$ which does not involve contact resistance. Next, at a fixed $\rho_{bulk}/\rho_{edge}$ ratio, the absolute resistance value $R_{1-2;5-6}$ scales linearly with the bulk (or, equivalently, edge) resistivity with a scale factor dependent on $\rho_{bulk}/\rho_{edge}$. Figure S 3 (d) plots the simulated ratio $R_{1-2;5-6}/\rho_{edge}$ (and also $R_{1-2;5-6}/\rho_{bulk}$, equal, of course, to $R_{1-2;5-6}/\rho_{edge}$ divided by $\rho_{bulk}/\rho_{edge}$) as a function of $\rho_{bulk}/\rho_{edge}$. This allows for independent evaluation of $\rho_{edge}$ and $\rho_{bulk}$ once $R_{1-2;5-6}$ and $\rho_{bulk}/\rho_{edge}$ are determined.

To reflect chiral nature of edge channels at finite B (although with a possibility of closely spaced counter-propagating edge states that would allow some back scattering), we experimented also with simulations of a very similar and only minimally-modified *diode-resistor* network, which blocks (either fully or only partially) current backflow along the edges (with the preferable direction controlled by the sign of *B*). Thus, unlike the linear all-resistor network, the non-linear diode-resistor network is inherently sensitive to the orientation of the applied magnetic field (or, equivalently, polarity of the feeding terminals), an observation of the model behavior that is fully supported by simulations. For the magnetic field pointing *into* the plane of the device in main text's Figure 1(b), current should prefer to flow clockwise along the outer edge of the device and counterclockwise along its inner edge. This is the case for the data measured at B=2 T shown in main text's Figure 2 (e)-(h). At high electron densities, i.e., deep in the IQHE regime, we simulate the experimental data using the *fully-blocking* diode-resistor network, while at the RP all-resistor network is more appropriate. For the lack of a detailed quantitative description of a crossover between these two regimes (where, intuitively, a partially-blocking diode-resistor network would be applicable), we attempted to simply patch them together along the crossover canyon boundary. Overall, the result is visually similar to the one obtained when processing data using the all-resistor network throughout [as depicted in main text's Figure 3 (c) and (d)]. Empirically, the edge in the diode-resistor network ends to be slightly more conductive (up to ~1.7x at largest $\rho_{bulk}/\rho_{edge}$) to compensate for the unidirectional current flow.

## IV. Transport with-edge versus no-edge

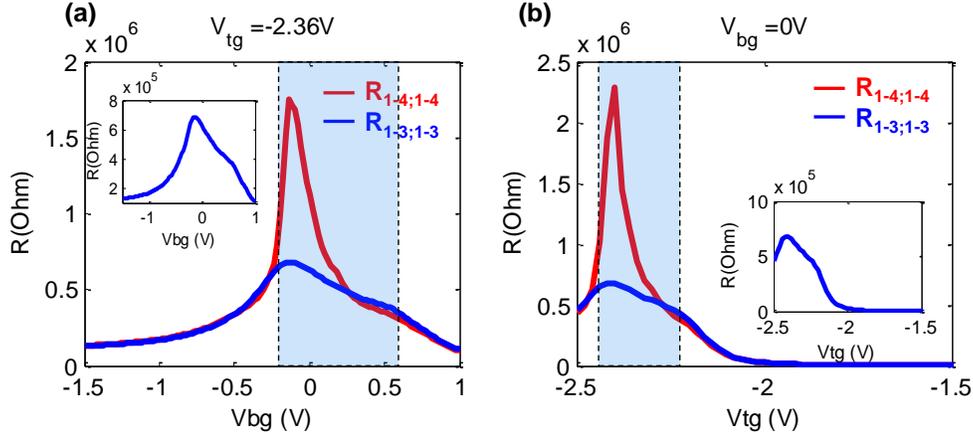

*Figure S 4. 1D line cuts of $R_{1-3;1-3}$ and $R_{1-4;1-4}$ as a function of a) back gate bias at $V_{tg}$=-2.36 V and b) top gate bias at $V_{bg}$=0V. The on-set of edge conductance is indicated by the kinks of the $R_{1-3;1-3}$ curves, as well as by the deviation of $R_{1-3;1-3}$ from $R_{1-4;1-4}$ (dashed boxes). When dominated by edge conductance, the negative slope of the $R_{1-3;1-3}$ curves suggests an electron-nature of the edge transport. The peak of $R_{1-3;1-3}$ of 7 kΩ corresponds to an edge resistivity of 1.6 kΩ/μm for 2 parallel edge channels with length of 880 μm each. This edge resistivity is close to the extraction value of 2 kΩ/μm achieved with the resistor network model.*

As discussed in the main text, the evidence of edge conductance can be seen from the comparison between $R_{1-3;1-3}$ (with edge) and $R_{1-4;1-4}$ (without edge). 1D line cuts of main text's Figure 1 (c) and (d) at a constant $V_{tg}$=-2.36 V and $V_{bg}$=0V are shown in Figure S 4 (a) and (b), respectively. When the bulk channel is highly conductive, there is no potential difference between leads 3 and 4 across the mesa, thus the two resistances overlap (outside of the dashed boxes). When the bulk becomes more resistive, leads 3 and 4 are electrically "disconnected", revealing a difference in conductance paths. Between leads 1 and 4, carrier must pass through the resistive bulk while between leads 1 and 3, carrier can travel along the edge if the bulk is too resistive. A deviation of $R_{1-3;1-3}$ from $R_{1-3;1-4}$ hence indicates the turn on of edge conductance.

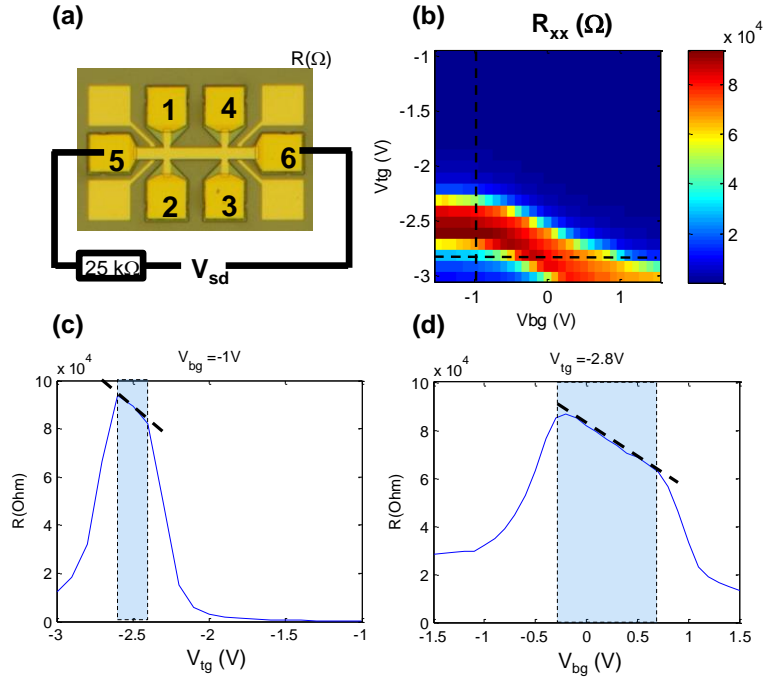

*Figure S 5 (a) schematic of a Hall bar , (b) Longitudinal resistance ($R_{xx} = V_{23}/I_{56}$) map as a function of top and back gate biases, dashed lines indicate biases where a 1D line cut is plotted in (c) Vbg=-1V and (d) Vtg=-2.8V. Dashed boxes in (c) and (d) highlight edge-transport dominated regime. The negative slope of the resistance curve suggests an electron-nature of the edge transport. The resistance maxima of 94 k$\Omega$ scales with the edge length (100 $\mu$m for 2 parallel edges) with a linear edge resistivity scale factor of ~1.9 k$\Omega$/$\mu$m. This edge resistivity is close to the extraction value of 2 k$\Omega$/$\mu$m achieved with resistor network model.*

Between leads 1 and 3, there are two possible parallel conductance path: (1) inside the bulk, along the mesa, and (2) at the outer edge of the mesa. The measured resistance is roughly the smaller between the two. When the bulk is more conductive than the edge (smaller resistance, outside of dashed boxes), the measured $R_{1-3;1-3}$ reflects the bulk resistance and when the bulk is more resistive than the edge (inside the dashed boxes), the measured $R_{1-3;1-3}$ reflects the edge resistance. The two kinks in $R_{1-3;1-3}$ curve right at the separation of $R_{1-3;1-3}$ and $R_{1-3;1-4}$ indicate a change in conductance mechanisms, which, in this case, is the transition from bulk dominant to edge dominant conductance. The same kinked character is also observed in a Hall bar device with parallel edge-bulk channel (*Figure S 5*). Note that in the Hall bar, the longitudinal resistance $R_{xx}$ does not involve contact resistance, so the kink-transition is not due to contact resistance effect. When dominated by the edge conductance, both $R_{1-3;1-3}$ of the Corbino ring and $R_{xx}$ of the Hall bar have a decreasing trend with positive gate bias, suggesting an n-type nature of electron transport. Assuming the measured edge resistance is linear with length, we estimate the resistivity per unit length of 1.6 k$\Omega$/μm and 2 k$\Omega$/μm for the Corbino ring and Hall bar, respectively. These values are close to the ~2 k$\Omega$/μm range numerically extracted from $R_{1-2;3-4}$ and $R_{1-2;5-6}$ data.

## V. Local versus non-local measurement

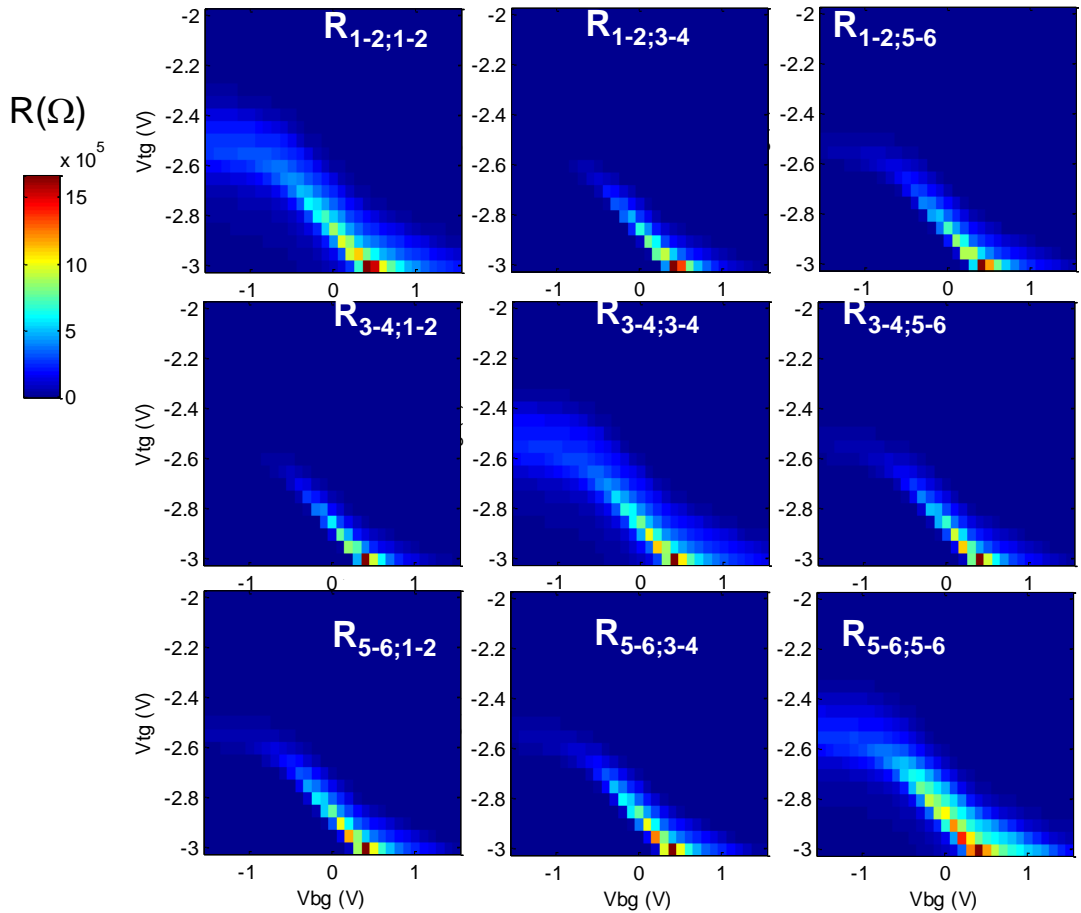

*Figure S 6 Local and non-local resistance maps under different measurement configurations.*

As a proof of edge conductance, Figure 2 in the main text already discussed non-local measurement with current running through leads 1-2 and voltage measured at leads 3-4 and 5-6. At the RP, voltages measured across 3-4 and across 5-6 are similar to the set voltage between 1-2. Same behavior is observed when current is running through the 3-4 pair or 5-6 pair and voltage measured across other pairs (*Figure S 6*). This clearly proves that the inner and outer edge are almost two equi-potential surfaces, a scenario only possible if there is substantial conductance along the edges.

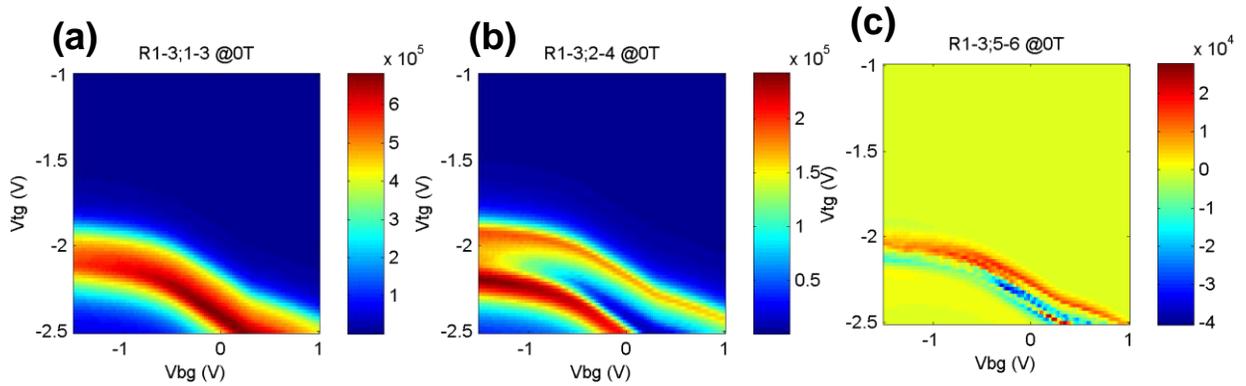

*Figure S 7 Quasi local measurement in the Corbino ring device suggesting the presence of edge-conductance: (a) $R_{1-3;1-3}$, (b) $R_{1-3;2-4}$ and (c) $R_{1-3;5-6}$.*

Evidence of edge conductance can also be seen from other quasi local measurement (e.g see *Figure S 7*). When current flows between 1-3, the discrepancy between local $R_{1-3;1-3}$ and quasi local $R_{1-3;2-4}$ delineates edge-dominated transport regime from bulk-dominated regime by the low resistance "tongue" in the middle of the resistance peak stripe of $R_{1-3;2-4}$. This is due to the electrical disconnection between 1 and 2, and between 3 and 4 when the bulk resistivity high, leads 2 and 4 become floating, resulting in low $R_{1-3;2-4}$. In addition, there is a substantial voltage built up between leads 5 and 6 at gate biases defining the resistance peak [*Figure S 7* (c)]. If the conductance is purely bulk-dominated, there should not be any built-up potential across 5-6, but it is not the case experimentally.

## VI. Measurement at B=1T

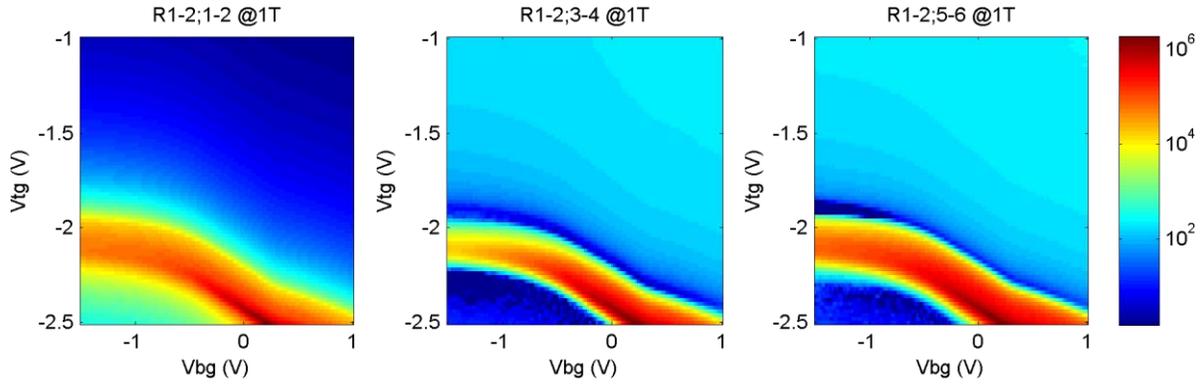

*Figure S 8 Local and non-local resistance maps under B=1T.*

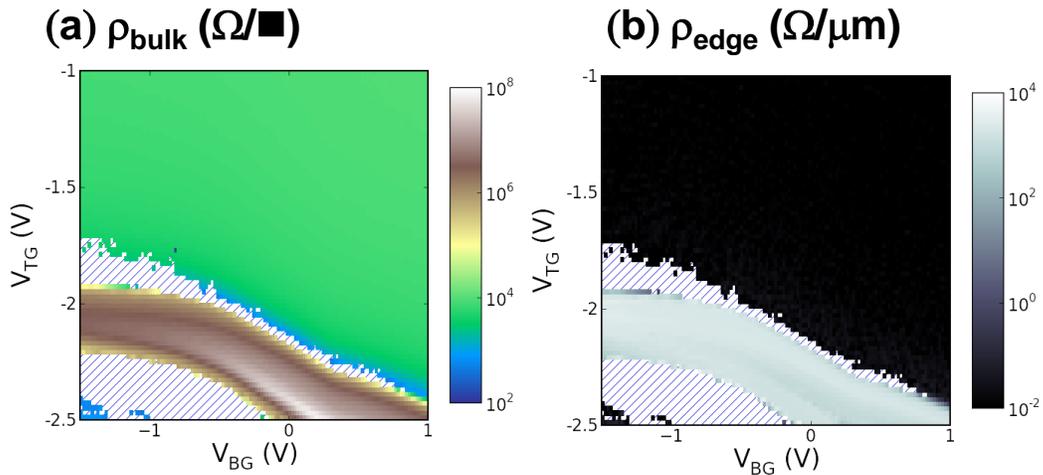

*Figure S 9 Extracted bulk and edge resistivity at B=1T*

Similar to analysis and extraction for B-0 and B=2T in the main text, additional data at B=1T are shown in *Figure S 8* and *Figure S 9* to illustrate the high dynamical range of the extraction technique. At B=1T, the ripplings are absent, but the increase of bulk resistance already allows for extraction of bulk and edge resistivity outside of the RP.

## Reference

1. Binh-Minh Nguyen, Wei Yi, Ramsey Noah, Jacob Thorp, and Marko Sokolich, Applied Physics Letters **106** (3), 032107 (2015).
2. Fanming Qu, Arjan J. A. Beukman, Stevan Nadj-Perge, Michael Wimmer, Binh-Minh Nguyen, Wei Yi, Jacob Thorp, Marko Sokolich, Andrey A. Kiselev, Michael J. Manfra, Charles M. Marcus, and Leo P. Kouwenhoven, Physical Review Letters **115** (3), 036803 (2015).
3. Lingjie Du, Ivan Knez, Gerard Sullivan, and Rui-Rui Du, Physical Review Letters **114** (9), 096802 (2015).
4. M. J. Manfra, E. H. Hwang, S. Das Sarma, L. N. Pfeiffer, K. W. West, and A. M. Sergent, Physical Review Letters **99** (23), 236402 (2007).
5. G. Chen, A. M. Hoang, S. Bogdanov, A. Haddadi, P. R. Bijjam, B.-M. Nguyen, and M. Razeghi, Applied Physics Letters **103** (3), 033512 (2013).